\documentclass[aps,prb,reprint,groupedaddress]{revtex4-1}

\usepackage{graphics}

\begin{document}


\title{\textit{In situ} determination of the anisotropy field in ferromagnetic films using magnetic susceptibility measurements by MOKE}

\author{A. Bartlett}

\author{R. Belanger}
\author{M. Amann}
\author{D. Venus}
\email{[corresponding author] venus@physics.mcmaster.ca}
\affiliation{Department of Physics and Astronomy, McMaster University, Hamilton, Ontario, Canada}

\date{\today}

\begin{abstract}
An alternate method of measuring anisotropy fields in thin film ferromagnets is demonstrated.  The method relies on the magnetic susceptibility in a small a.c. magnetic field, measured \textit{in situ} using the magneto-optic Kerr effect (MOKE), and will be useful in situations where more specialized apparatus are not conveniently available, or constraints discourage the use of a large, static magnetic field.  The method is demonstrated for Co/W(110) films, where it yields anisotropy fields in agreement with previous studies using more conventional torque magnetometry.  The sensitivity of the method is demonstrated using CoO/Co/W(110) bilayer films, where the anisotropy due to interfacial exchange coupling is detected and used to find the N\'{e}el temperature of the thin CoO layer.
\end{abstract}


\maketitle

\section{Introduction}
A central part of research in ferromagnets involves the determination of effective fields, such as the magnetic anisotropy,  internal to a ferromagnetic system.  Effective fields are a reflection of the structural properties (such as the lattice, electronic structure,  grain or sample shape and dimensions) that add to the great variety of properties across a spectrum of magnetic systems.   In thin magnetic films and multilayers, the effective fields introduced by surfaces and interfaces can result in novel magnetic characteristics through, for example, surface anisotropies, localized strain fields, interlayer coupling and periodic defect structures\cite{Sander}.  In order to understand and exploit this potential, a variety of experimental methods for measuring internal effective fields have been adapted to  thin film geometries.  Among these are the classic methods of torque magnetometry\cite{Gradmann}, Brillioun light scattering\cite{BLS}, and ferromagnetic resonance (FMR)\cite{FMR}, all of which determine the effective fields by measuring a very small angular deflection of the magnetization vector in a transverse field.

Although these classic methods set the standard for measuring anisotropies in thin films, most researchers make limited use of them because of practical considerations, including lack of access to a specialized, dedicated apparatus, the need for an \textit{in situ} technique compatible with ultrahigh vacuum conditions for thin film studies, and a desire to avoid using a large, static external magnetic field that may compromise the subsequent use of electron spectroscopies.  As a result, a large portion of the research community has turned instead to relatively simple, widely available \textit{in situ} magneto-optical Kerr effect (MOKE) techniques\cite{Qiu} for routine characterization.  Since domain processes are dominant in easy axis hysteresis loops, the anisotropy is usually determined from hard axis hysteresis loops or the approach to saturation in hard axis magnetization curves\cite{Hajjar}.  This requires an applied static field of the order of the anisotropy field and thus remains a large field technique.   In some thin film systems it is possible to find conditions where applying a static transverse field delays the onset of domain processes in (nearly) easy axis hysteresis loops.\cite{Weber,Leeb}  Then the anisotropy can be extracted from the linear, low field region of the loop.

The present article illustrates an \textit{in situ} MOKE technique to determine anisotropy fields in thin films that combines aspects of both these methods.  It uses hard axis measurements, but in a field that is so small that domain processes are not activated.  That is, the dynamic magnetic susceptibility is measured in a small a.c. field, $H_{app}$, applied transverse to the easy axis, so that a linear deflection of the magnetization vector is measured.  Transverse susceptibility measurements have been used to determine anisotropies in thick, bulk-like films\cite{Zimmermann}, and the development of micromechanical resonators for susceptibility measurements and magnetometry is extending the sensitivity limits for the characterization of submonlayer depositions,\cite{Min} and individual mesoscopic magnetic elements.\cite{Losby}   However, to our knowledge, \textit{in situ} MOKE susceptibility measurements have not been applied to the routine characterization of magnetic anisotropies.  This is likely because the measurement of the deflection of the magnetization in a small field of order 0.1 kA/m requires detection of a Kerr rotation of order
\begin{equation}
\label{envelope}
\phi \sim \frac{M_\perp}{M_S} \Phi_{cal} \sim \frac{H_{app}}{H_{anis}} \Phi_{cal} \sim 10^{-8} \  \mathrm{rad/ML} ,
\end{equation}
where $\Phi_{cal}$ is the calibrated Kerr rotation for magnetic reversal per ML ferromagnetic film.  Although the measurement of absolute Kerr angles of order 10 nrad might be considered challenging, the present article shows that this is a feasible and reliable method that can be employed when more traditional techniques are inappropriate or unavailable.

The method is illustrated using two applications.  First, a  determination of the in-plane anisotropy field at room temperature in Co/W(110) films is compared quantitatively to a previous study\cite{Fritzsche} using torque magnetometry.   It is then generalized in a straightforward manner to characterize the anisotropy as a function of temperature.  The method is similarly amenable to measurements as the ultrathin film is being grown.  Second, when applied to a CoO/Co/W(110) bilayer film, the temperature-dependent effective field reveals the anisotropy due to interface exchange coupling between the antiferromagnet and ferromagnetic layers.  This is used to infer the N\'{e}el temperature of the thin antiferromagnet film.

%


\section{Experimental methods}
The magnetic anisotropy in a ferromagnet is directly related to the magnetic susceptibility measured using a small applied transverse field.  The effective field\cite{MacDonald} internal to the magnetic system, $\mathbf{H}_{eff}$, is given by the sum of the field applied externally $\mathbf{H}_{app}$ and the the anisotropy field $\mathbf{H}_{anis}$,
\begin{equation}
\label{Heff}
\mathbf{H}_{eff}=\mathbf{H}_{app}+\mathbf{H}_{anis}.
\end{equation}
The anisotropy field is related to the anisotropy energy density, $E_{anis}(\mathbf{M})$as
\begin{equation}
\label{anis energy}
 \mathbf{H}_{anis}(\mathbf{M})=-\frac{1}{\mu_0}\frac{\partial E_{anis}(\mathbf{M})}{\partial \mathbf{M}},
\end{equation}
where the anisotropy energy includes the dipole-induced shape anisotropy as necessary.  Taking the derivative of eq.(\ref{Heff}) with respect to $\mathbf{M}$, and re-arranging gives diagonal components of the susceptibility tensor
\begin{equation}
\label{chitensor}
\chi_{ii}=\frac{\partial M_i}{\partial H_{app,i}}=\frac{\chi_{eff,ii}}{1+\frac{1}{\mu_0}\frac{\partial^2 E_{anis}(\mathbf{M})}{\partial M_i^2}\chi_{eff,ii}},
\end{equation}
where the index $i$ refers to a Cartesian axis.  Far from a Curie transition, $|\mathbf{M}|=M_S$ does not change quickly, and, unless the anisotropy is extremely weak, the factor of one in the denominator can be neglected.
\begin{equation}
\label{chi}
\chi_{ii}^{-1}=\frac{1}{\mu_0}\frac{\partial^2 E_{anis}(\mathbf{M})}{\partial M_i^2}.
\end{equation}

\begin{figure}
\scalebox{.35}{\includegraphics{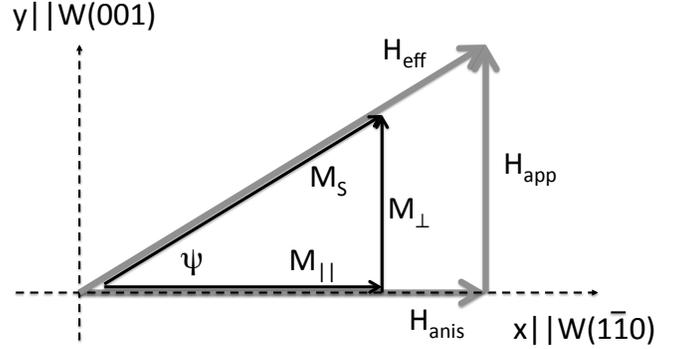}}
\caption{\label{fig1}In an in-plane ferromagnet, the anisotropy field is aligned with the easy x-axis.  Applying an external field along the y-axis creates a net effective field at angle $\psi$ (greatly exaggerated) to the easy axis.  Because the saturation magnetization lies along the effective field, the magnetization components (black lines) form a triangle that is similar to that formed by the magnetic fields (grey lines).}
\end{figure}

This relation is illustrated concretely in fig.(\ref{fig1}) for the ferromagnetic Co/W(110) films in these experiments.  The anisotropy field lies along the in-plane easy x-axis where the anisotropy energy is minimized.  An external field is applied in-plane, but transverse to the anisotropy field, so that the effective field makes an angle $\psi$ with the easy axis.  The magnetic energy is minimized when the saturation magnetization $M_S$ lies along $H_{eff}$, creating a similar triangle of sides $M_{\perp}$ and $M_{\parallel}$.  

Co films grow epitaxially on the b.c.c. W(110) substrate in a strained cubic structure\cite{Knoppe,Fritzsche}.  The first 2 nm have constant strain along the W(001) axis, and the strain relaxes with increasing thickness for thicker films.  Because of the in-plane magnetization geometry, there is no demagnetization field.  The rectangular lattice of the W(110) surface and the uniaxial strain axis allow a 2-fold surface anisotropy $K_{2s}$, and the uniformly, uniaxially strained region creates a bulk-like anisotropy $K_2$.  The cubic structure of the volume of the Co film creates a 4-fold bulk anisotropy $K_4$ and the depth-dependent strain relaxation in the thicker films cannot be distinguished from a surface anisotropy and contributes to $K_{4s}$.  The anisotropy energy of the film is then given by   
\begin{equation}
\label{Co anis}
E_{anis}=(K_2+K_{2s}/t) \sin^2\psi + (K_4+ K_{4s}/t) \sin^2\psi \cos^2\psi.
\end{equation}  
Since $M_\perp =M_y=M_S \sin\psi$, eq.(\ref{chi}) yields 
\begin{equation}
\label{Kdef}
\frac{M_S}{\chi_{yy}}=\frac{2(K_2 + K_4)+2(K_{2s} + K_{4s})/t}{\mu_0 M_S}\equiv\frac{2(K+K_s /t)}{\mu_0 M_S}.
\end{equation}

For a small applied field, the similar triangles in fig.(\ref{fig1}) also show that
\begin{equation}
\label{anis H}
H_{anis}=\frac{H_{app}}{ \tan\psi} \approx \frac{M_S}{M_{\perp}/H_{app}}=\frac{M_S}{\chi_{yy}},
\end{equation}
where the small angle approximation $\tan\psi \approx \sin\psi$ has been used.  The equivalence of eq.(\ref{Kdef}) and (\ref{anis H}) shows that the surface and volume anisotropy constants can be isolated by plotting
\begin{equation}
\label{plot}
\mu_0 H_{anis} \, M_S\, t = 2Kt + 2K_s 
\end{equation}
as a function of $t$.

The ratio of the magnetization components $M_S$ and $M_{\perp}$ in eq.(\ref{anis H}) can be measured using the magneto-optical Kerr effect (MOKE)\cite{Qiu}.    When plane s- or p-polarized light is incident on a magnetized sample, the plane of polarization of the reflected light is rotated by a Kerr angle $\phi$ with respect to that of the incident light, and the rotation is proportional to both the thickness of the ferromagnetic film and the magnitude of the magnetization vector.  In the present experiments, the longitudinal Kerr effect is utilized -- that is, the plane of  incidence of the light contains the in-plane magnetization component being detected, and the surface normal.  

The apparatus used in the present measurements is described in detail in ref.(\onlinecite{Arnold2}) and (\onlinecite{Arnold4}).  Light from a HeNe laser passes through a UHV window, and strikes the sample at an angle of $45^o$ from normal incidence.  The reflected light exits the vacuum chamber through a second window, passes through a second polarizing crystal and is absorbed by a photodiode.  The polarizing crystals themselves should have an extinction ratio when crossed of $<10^{-6}$, and the polarizer mounts should be capable of fine angular adjustments of order arcminutes.  For almost crossed polarizers, changes in the Kerr rotation create changes in the intensity at the photodiode.  The axis of a Helmholtz pair of field coils are aligned with the surface of the crystal and the sample may be rotated so that any in-plane direction lies along this axis.  The coils are used to produce a small a.c. field, or a larger field pulse, but tend to overheat if used to make the sustained d.c. field needed for a standard hysteresis loop.  The entire vacuum chamber is at the centre of three large, orthogonal Helmholtz pairs of coils used to create a field free region at the sample position.

Detection of very small Kerr rotations requires passing linearly polarized light through polarizing crystals that are very nearly crossed.  For this reason, a compensation method is used to remove ellipticity induced by the windows in the vacuum chamber.\cite{Arnold4}  Since the polarizers and the compensation alignment are imperfect, a fraction $\epsilon$ of the incident light passes through the crossed polarizers.  For small relative rotations $\theta$ of the polarizers from the crossed condition, the light intensity passing through the polarizers and detected by a photodiode is
\begin{equation}
\label{extinct}
I(\theta)=I_0[\theta^2 +\epsilon].
\end{equation}
The value of $\epsilon$ can be determined for any optical alignment by finding the angle $\theta_\epsilon$ where the intensity is doubled from the crossed condition, $I(\theta_\epsilon )=2I(0)$.  Then
\begin{equation}
\label{epsilon}
\epsilon=\theta_\epsilon^2.
\end{equation}
This relation allows absolute calibration of the optical rotation. Setting the polarizers at a relative angle $\theta_{set}$ gives a differential change in the intensity $dI$ due to the small Kerr rotation $d \theta \equiv \phi$, so that 
\begin{equation}
\label{dI}
\phi=\frac{dI}{2I(\theta_{set})} \frac{\theta_{set}^2 + \theta_\epsilon^2}{\theta_{set}}.
\end{equation}
\begin{figure}
\scalebox{.45}{\includegraphics{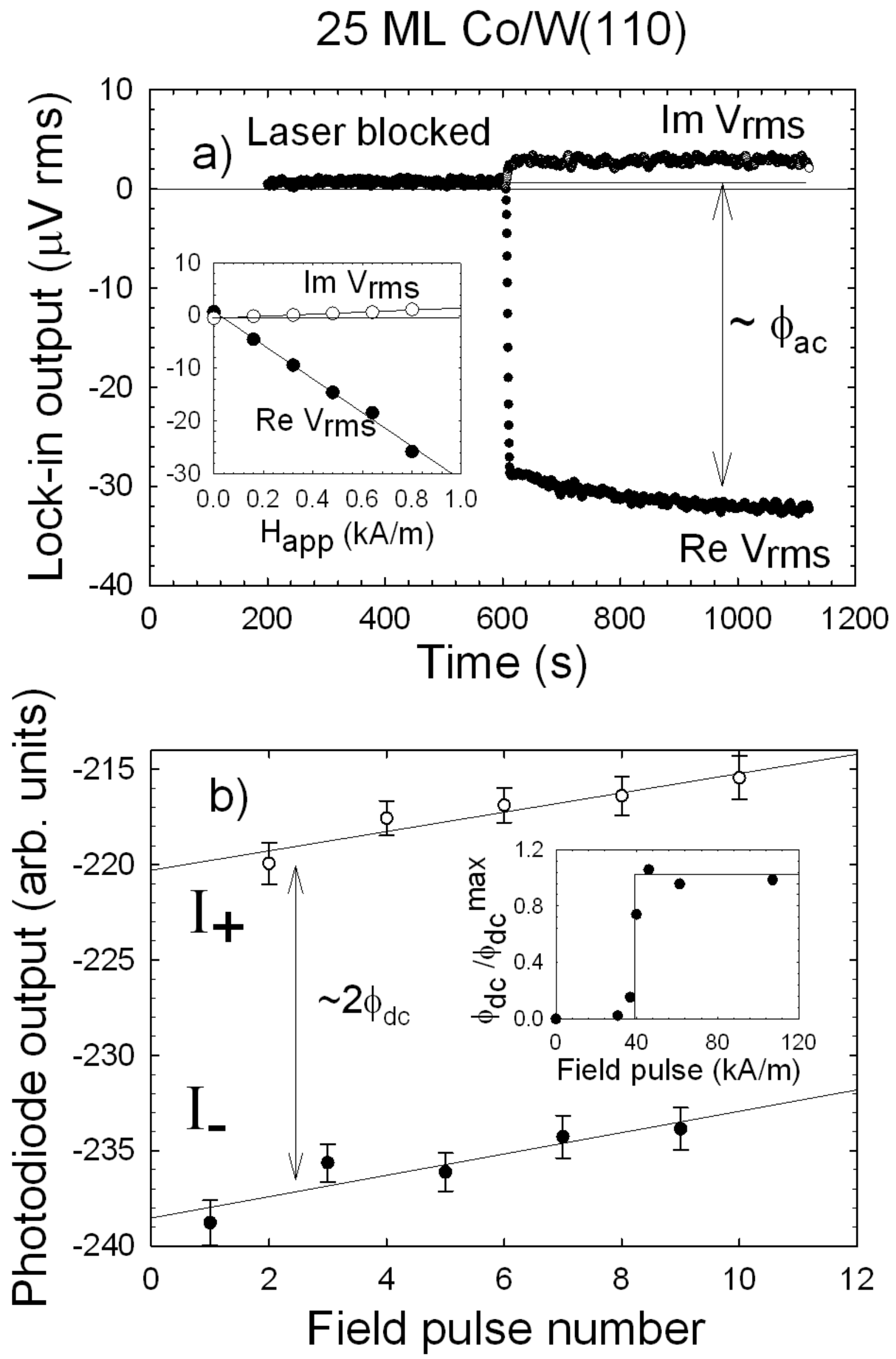}}
\caption{\label{fig2}Measurement of the a.c. and d.c. Kerr rotations for a 25 ML Co/W(110) film.  a) A transverse a.c. field of 0.80 kA/m rms at 210 Hz induces a small transverse component of the magnetization that produces an a.c. Kerr rotation in the longitudinal Kerr effect geometry.  The photodiode signal is input to a lock-in amplifier that gives an rms voltage in-phase (Re) and out-of-phase (Im) with the field.  The insert shows that the magnetic response is linear at least up to an applied field of 0.80 kA/m rms.  The units on the vertical axis are the same as in the main figure.  b)  A field pulse is applied along the easy axis of the ferromagnet to reverse the magnetization.  The d.c. output of the photodiode moves between the levels $I_+$ and $I_-$ as the pulse polarity is reversed.  The inset shows the relative change in the d.c. Kerr angle in remanence as a function of the pulse field amplitude. The response is close to an ideal step response, indicating a square hysteresis loop.}
\end{figure}

To determine $M_\perp$, the plane of light scatting is the yz plane of fig.(\ref{fig1}), and the field coils generate a small a.c. field $H_{app}$ along the y axis, at a frequency of 210 Hz.  The pre-amplified photodiode signal is fed into a lock-in amplifier along with a reference signal derived from the current supplied to the field coils.  Optimizing the signal-to-noise depends upon noise components generated by the laser stability, the mechanical stability of the sample, and by the photodetector, that are specific to the apparatus.\cite{Arnold2}  For our setup, $\theta_{set}=24$ arcminutes is used in this a.c. mode.  The d.c. photodiode output at this angle measures $I(\theta_{set})$. The output of the lock-in amplifier, as illustrated in fig.(\ref{fig2}a),  is an rms voltage that measures $dI$.  Both Real (in-phase with the a.c. field) and Imaginary (out of phase with the a.c. field) responses are provided by the lock-in.  $\phi_{ac}$ can then be calculated from eq.(\ref{dI}).

The inset to fig.(\ref{fig2}a) shows the response of the magnetization as a function of the transverse a.c. field strength.  The units on the vertical scale are the same as in  the main figure.  The response is linear up to at least 0.8 kA/m rms.  The fact that the imaginary component of the response is very small confirms that the magnetization undergoes a reversible angular deflection, and that domain wall processes are not important.  All the data presented in this article were measured with an a.c. field of 0.8 kA/m rms.

As can be seen in the figure, there is small background signal when the field is applied, but the laser is blocked.  This has been minimized by attention to ground loops.  An independent background is also observed in the inset when the a.c field is zero (but the reference signal is still present at the lock-in) and the laser is not blocked.  This measures the background at the measurement frequency, and is mostly due to mechanical motions of the apparatus.  Accounting for both of these backgrounds can be important for measurements of order nrad.

To determine $M_S$, the applied field and scattering plane of the light is rotated $90^o$ from the orientation in fig.(\ref{fig1}), so that $H_{app}$ lies along the x-axis.  A large capacitor is discharged through the Helmholtz coils to generate a field pulse of about 35 ms with a maximum strength of $\sim$ 110 kA/m.    The inset to fig. (\ref{fig2}b) shows the variation of the d.c. Kerr rotation in remanence as a function of the strength of the field pulse.  The data points lie very close to the idealized step expected for a square hysteresis loop that switches a single crystal sample by domain wall propagation.  They are inconsistent with the gradual change in remanence that traversing minor loops of an ``S"-shaped hysteresis loop would produce.  This indicates that the change in d.c. Kerr angle is, to a very good approximation, that corresponding to a magnetization change of $2M_S$.

The d.c. photodiode signals $I_+$ and $I_-$ are measured between reversal pulses, as is illustrated in the main part of fig.(\ref{fig2}b).   For d.c. measurements, the signal-to-noise is optimized\cite{Arnold2} by the choice $\theta_{set}=\theta_\epsilon$, so that eq.(\ref{dI}) gives the optical rotation corresponding to $M_S$ as
\begin{equation}
\label{dc}
\phi_{dc}=\frac{I_+ - I_-}{I_+ + I_-}\ \theta_\epsilon.
\end{equation}
For routine measurements, the extinction ratio of the entire optical setup is typically $\epsilon \le 10^{-5}$.  Although lower values giving a higher signal-to-noise ratio are achievable through careful optical alignment and the use of apertures\cite{Arnold4}, the alignment is then more sensitive to small perturbations that cause a drift in the average d.c. background.   Slow drift rates can be achieved through rigid construction, with optical rails bolted directly to the UHV windows and a relatively short manipulator shaft on the sample holder.  
The remaining drift is evident in the figure, and is easily accounted for through repeated measurement.  Given that a measurement for a single field pulse is an average over about 5 minutes, the background drift rate in the present measurements is of order 1 part in 2000 per minute.
 
Because the a.c. and d.c. Kerr rotations are measured from the same film in a longitudinal geometry, and are proportional to the film thickness, the anisotropy field in eq.(\ref{anis H}) can be written in the internally calibrated manner
\begin{equation}
\label{Hanis}
H_{anis}=\frac{\phi_{dc}}{\phi_{ac}} H_{app}.
\end{equation}

\section{Results and analysis}
\subsection{In-plane anisotropy of Co/W(110) films}
A first demonstration of the method is to measure the in-plane uniaxial magnetic anisotropy of Co films on W(110), a system for which torque magnetometry results can provide a benchmark.\cite{Fritzsche}  The films were grown on a W(110) single crystal substrate, the cleanliness and structure of which were confirmed using Auger electron spectroscopy (AES) and low energy electron diffraction (LEED).  

The deposition for the thermally stable\cite{Garreau} first Co ML was calibrated by attenuation of the W AES signal as a function of Co deposition.  This showed a clear break between linear segments when the film was annealed to 600 K after each sequential deposition.  Each Co film was grown at room temperature in two steps, with annealing to 600 K after the first ML to promote wetting, and then annealing the full thickness to 400K (which is below the temperature at which magnetic properties such as the critical temperature are affected\cite{Garreau}) to ensure stability under temperature cycling.  The LEED pattern showed the well-documented\cite{Fritzsche,Knoppe} satellite spots indicating strained epitaxy along the W[001] in-plane direction for thickness less than about 10 ML.  The pattern became indistinguishable from hexagonal for thickness above about 10 ML.  The Co interlayer spacing was measured using the periodic intensity maxima of the specular LEED spot, and found to be $2.02\pm0.09$ \AA, in agreement with half the c-axis lattice parameter of h.c.p. Co.  In the absence of a high resolution LEED apparatus, no detailed characterization of the strain evolution of the film growth was made.  

The in-plane easy axis of magnetization of the Co film was confirmed to be along W[1$\overline{1}$0]  by measurements such as those shown in fig.(\ref{fig2}).  In order to establish a consistent magnetic history, each film was subjected to ten magnetization reversals by field pulses before any MOKE measurements were made.

\begin{figure}
\scalebox{.5}{\includegraphics{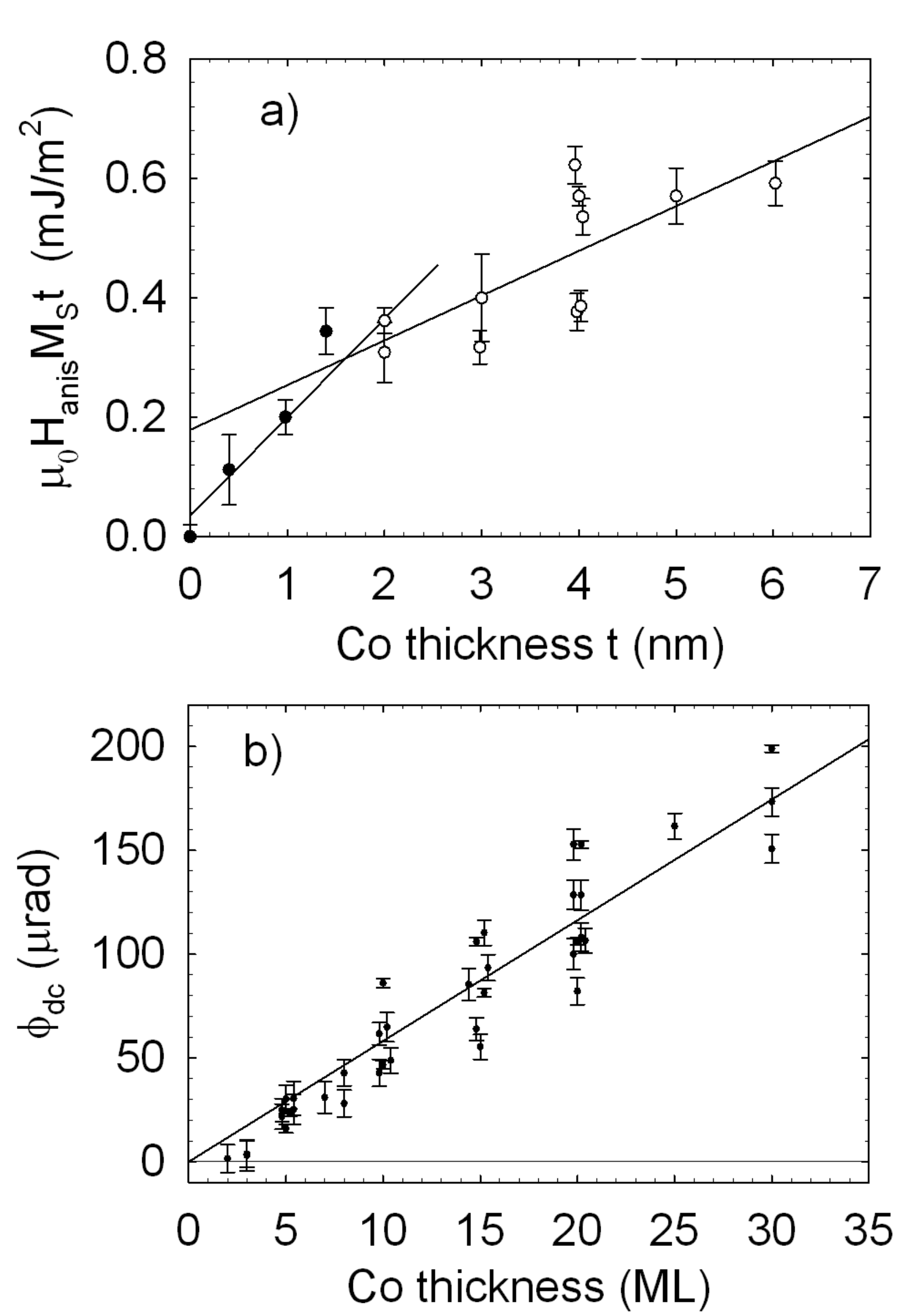} }
\caption{\label{fig3} a) The surface anisotropy energy density is plotted against the Co film thickness for a series of independently grown films.  Following ref.(\onlinecite{Fritzsche}), the data is separated into two parts.  For thickness $\le$ 2 nm (solid symbols) the film grows with a strained structure on W(110).  For thickness $\ge$ 2 nm (open symbols), the strain relaxes.  For each part, the slope of the fitted line gives the volume anisotropy, and the intercept gives the surface anisotropy.  b) The dc Kerr rotation is plotted as a function of Co film thickness.  The slope gives the calibration constant $\Phi_{cal}=5.8\pm0.3 \; \mu$rad/ML. }
\end{figure}

Measurements of the anisotropy field of a series of independently grown films of different thickness are presented in fig.(\ref{fig3}a).  For this plot, the saturation value of bulk Co, $\mu_0 M_s$= 1.82 T, is used.\cite{Chikazumi,Fritzsche}  Following ref.(\onlinecite{Fritzsche}), the data is separated into two parts.  For thickness $t \ge 2$ nm (open symbols), the strain in the cubic structure relaxes with increasing thickness. In this range, the slope and intercept of a least-squares linear fit gives volume and surface contributions of $2K = (0.81\pm0.25)\times 10^5$ J/m$^3$, and $2K_s = 0.19\pm0.09$ mJ/m$^2$, respectively.  The volume anisotropy of these thicker films is in good agreement with the earlier results\cite{Fritzsche} (which do not, however, have a quoted estimate of uncertainty).  The surface anisotropy, however, is a factor of 3 smaller ($K_s = 0.095\pm0.045$ mJ/m$^2$ in the present study vs. $0.3$ mJ/m$^2$).   For thickness less than 2 nm (solid symbols) the film grows with a consistently strained structure. In this range, the fitted line yields $2K = (1.9\pm0.4)\times 10^5$ J/m$^3$, and $2K_s = 0.02\pm0.04$ mJ/m$^2$.  The null result for the surface anisotropy is in agreement with ref.(\onlinecite{Fritzsche}) and the volume anisotropy is again about 3 times smaller in the present work.

These differences may be attributable to differences in the preparation of the films.  The films in the present study were annealed to 600K after 1 ML Co deposition to encourage wetting of the substrate, and then to 400K after the complete deposition to ensure stability during temperature cycling.  In the present study, films at all thicknesses yielded square loops, as can be seen by the consistency of the linear dependence of $\phi_{dc}$ in fig.(\ref{fig3}b).  In ref.(\onlinecite{Fritzsche}), the film growth and the measurements were performed at room temperature with no annealing. Their report that the first ML is magnetically ``dead"  may indicate incomplete wetting, so that the density of surface defects is increased.  This would be consistent with the S-shaped hysteresis loops they report for thinner films, that could not be saturated.  (See, for example, fig.6b for a 7.9 ML Co film in ref.\onlinecite{Fritzsche})   If the annealing procedures in the present study relieved the strain in the films to some degree compared to the previous study, then the strain contributions to $K_{2}$ for $t \le$ 2 nm and to $K_{4s}$ for $t \ge$ 2 nm would be reduced by an equal proportion.   While this provides a consistent explanation of the differences, without a high resolution LEED we cannot quantify possible differences in the strain evolution between the films prepared using different procedures.

The measurements of $\phi_{dc}$ for the films in fig.(\ref{fig3}a), and for other films measured during the preliminary parts of the study, are collected in fig.(\ref{fig3}b).  This provides a calibration of longitudinal MOKE for the Co/W(110) system, with the fitted line giving $\Phi_{cal}=5.8\pm0.3 \; \mu$rad/ML.  The scatter in fig.(\ref{fig3}b) suggests that the major source of error in using small-field susceptibility measurements to determine magnetic anisotropies is the reproducibility of a series of independently-grown films.  This is a common theme for many ultrathin films studies.
\begin{figure}
\scalebox{.5}{\includegraphics{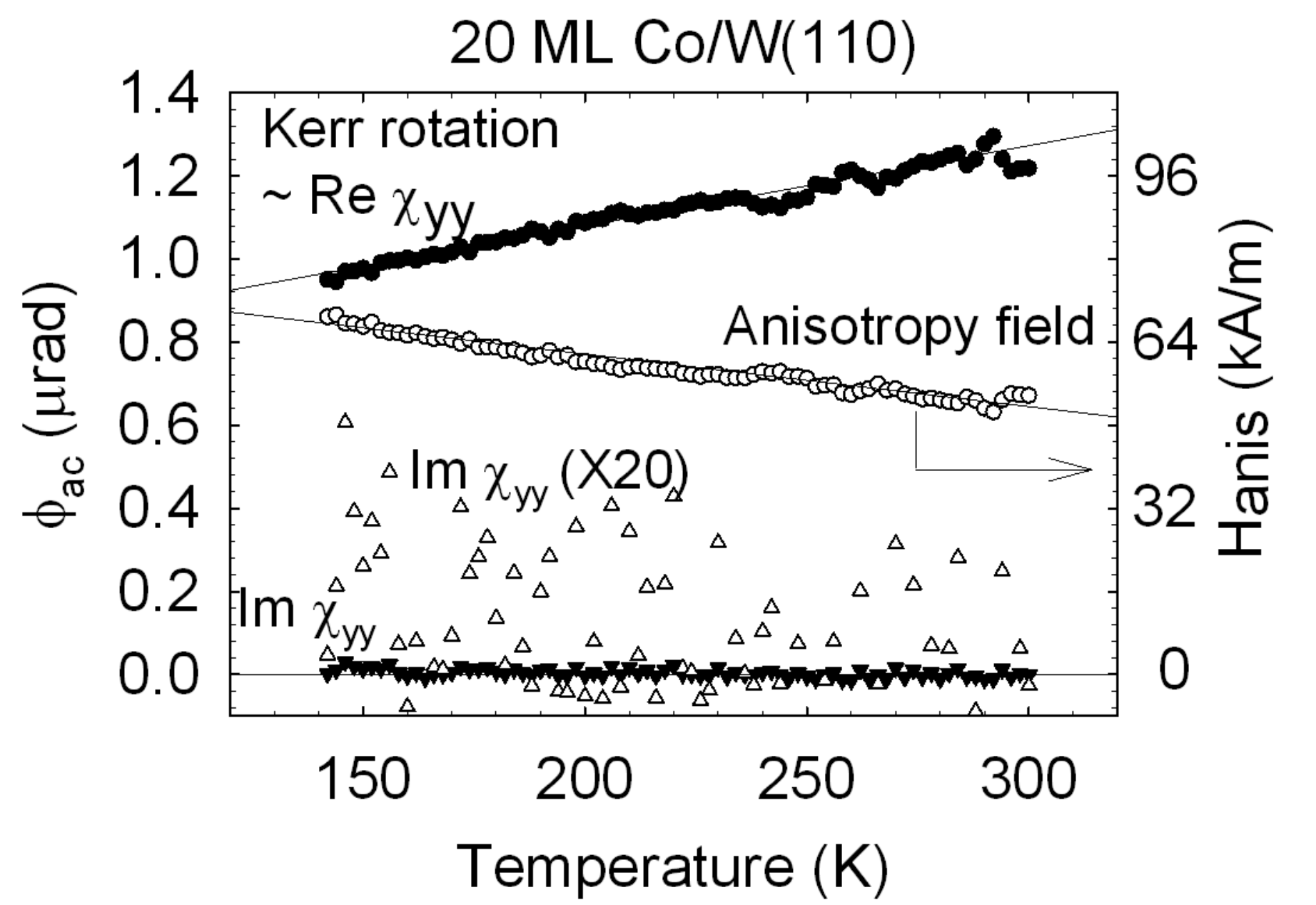}}
\caption{\label{fig4}The Re and Im parts of the a.c. Kerr rotation are plotted, using solid symbols, as a function of temperature as the film is heated.  These are proportional to Re$\chi_{yy}$ and Im$\chi_{yy}$.  Im$\chi_{yy}$ is also shown magnified by 20 times (open triangles), revealing no structure beyond noise.  The anisotropy field derived using eq.(\ref{Hanis}) is plotted using open circles.}
\end{figure}

The temperature dependence of the magnetic anisotropy of the Co/W(110)  films can be measured using a straightforward extension of the method.  An example is shown in fig.(\ref{fig4}).  The sample was prepared as before, then cooled.  The optical alignment was performed when cooled, and then measurements recorded $\phi_{ac}$ in a small transverse field as the sample temperature was increased at an average rate of 0.02 K/s.  Given the long measurement times, $I(\theta_{set})$ in eq.(\ref{dI}) was also recorded continuously, so that small drifts in the alignment could be properly accounted for.  The data were binned in 2 K increments before plotting.

 $\phi_{ac}$, which is proportional to Re$\chi_{yy}$, decreases linearly as the  temperature decreases. Im$\chi_{yy}$ is very small, and shows no structure, even when magnified by a factor of 20.  This confirms that the magnetic response is a reversible rotation unrelated to dissipative mechanisms.  The anisotropy field derived from eq.(\ref{Hanis}) decreases linearly with increasing temperature, with a linear coefficient $\lambda/K(T=0)= -(1.25\pm0.03)\times10^{-3}$ K$^{-1}$. 
The 4-fold volume anisotropy of a cubic material is usually observed to decrease as a higher power of the magnetization\cite{Chikazumi}, $M_S(T)/M_S(T=0)$.  However, the temperature dependence of strain-induced anisotropies, whether surface- or bulk-like, are instead related to the temperature dependence of bulk elastic and magnetostrictive constants. For example, FMR measurements on Ni/W(110) films shown a linear temperature dependence of the 2-fold surface anisotropy due to strain relaxation of thicker films.  This can be explained by the temperature variation of the bulk elastic and magnetostrictive constants,\cite{Farle} with $\lambda/K(T=0)$ of order $10^{-3}$ K$^{-1}$ .  Similar measurements for Ni/Cu(001) show a linear temperature dependence of 2-fold bulk anisotropy due to the uniform strain in these films.  The present results for Co/W(110) are difficult to interpret unambiguously, as they represent the temperature dependence of the sum of strain and cubic crystalline contributions to the anisotropy.  However, since anisotropies due to both regions of uniform strain and regions of relaxing strain play an important role in the surface and bulk anisotropies in this system, it is reasonable that they are also prominent in the temperature dependence of the total anisotropy.

\subsection{Anisotropy due to layer exchange in CoO/Co/W(110)}
A second demonstration of the method is to measure the effect of the exchange coupling of the spins at the interface between an antiferromagnet (AFM) and a ferromagnet (FM), on the anisotropy field of the ferromagnet\cite{Berowitz}.  AFM/FM interfacial coupling, or the phenomenon of unidirectional interfacial exchange bias, is of great importance in memory, sensor and other devices based upon magnetic multilayers.
\begin{figure}
\scalebox{.5}{\includegraphics{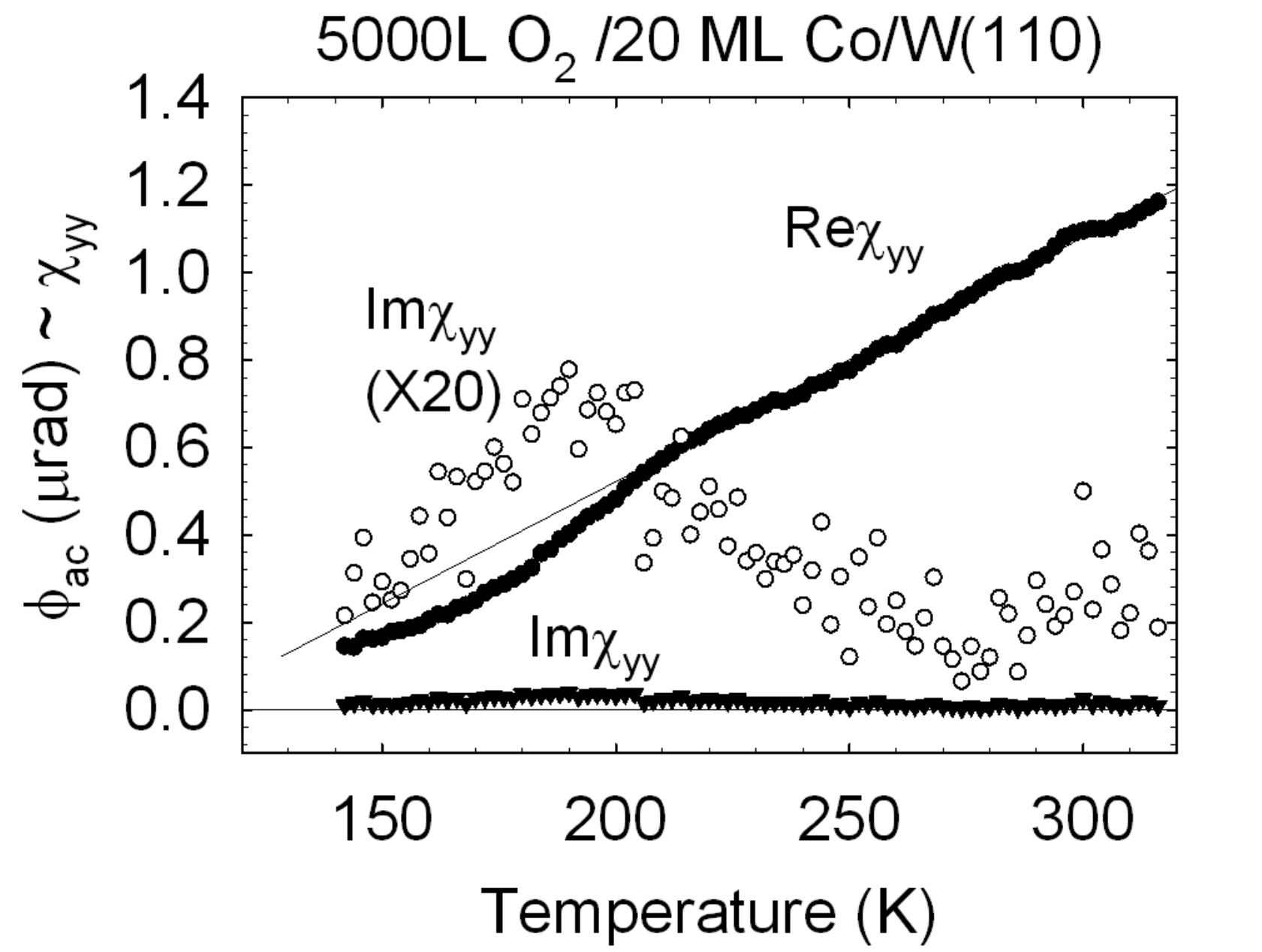}}
\caption{\label{fig5}The Re and Im parts of the a.c. Kerr rotation are plotted in closed symbols for a Co/W(110) sample that has been exposed to 5000L O$_2$, so that a polycrystalline CoO layer forms.  Im$\chi_{yy}$ is also shown magnified 20 times using open symbols.  It exhibits a broad peak, indicating dissipation in the temperature range where  Re$\chi_{yy}$ departs from a linear temperature dependence.}
\end{figure}

The conventional method of detecting the presence of the AFM/FM coupling is through the shifting of the FM hysteresis loop, so that is not centred on the field axis.\cite{Nogues}  In order to observe this exchange bias, the bilayer must be field-cooled through the N\'{e}el temperature, $T_N$, of the AFM in order to define a preferred direction along the axis of magnetic anisotropy.  Of course, this shift disappears once the bilayer is heated above $T_N$.  However, even without a field-cooling procedure, the change in the anisotropy field of the FM should be detectable by measurements of the susceptibility in a small field transverse to the anisotropy axis, as transverse deflections of the magnetization do not depend on whether the anisotropy is unidirectional or uniaxial.

Following previous studies\cite{Gruyters,Clemens,Klingenberg} of the formation of CoO, CoO/Co/W(110) bilayers were prepared by exposing a 20 ML Co film to 5000 L of O$_2$ at room temperature.  The LEED patterns of these films show a hexagonal pattern of rather diffuse spots with an in-plane spacing of $3.0\pm0.1$ \AA.  This is consistent with the spacing expected from the bulk CoO lattice, but is not conclusive because of the large uncertainty due to the diffuse spots.  However, a LEED analysis yielded an interlayer spacing of $2.41\pm0.09$ \AA, in good agreement with the value of 2.459 \AA $\;$expected from bulk CoO.  Since these results are in agreement with previous studies, and inconsistent with the previously-determined layer spacings of Co films and of the W(110) substrate, it was concluded that a preferentially-oriented, but polycrystalline, layer of CoO had been formed on the Co.  The relative thicknesses of the CoO and the remaining Co is not clear, but the CoO film is at least as thick as the probing layer of the LEED electrons in the range of 25 to 300 eV.

Fig.(\ref{fig5}) presents a measurement of the temperature-dependent Kerr rotation of the bilayer, $\phi_{ac}$, which is proportional to the susceptibility in a small transverse field, $\chi_{yy}$.  The qualitative changes due to formation of a layer of CoO can be seen by comparing to fig.(\ref{fig4}).  The susceptibility again decreases linearly with decreasing temperature, but now only to about 200K.  At this temperature, the susceptibility falls below the linear trend, and Im$\chi_{yy}$ shows a clear, broad peak that indicates dissipation.  These features are consistent with a change in the anisotropy field due to the formation of interfacial exchange coupling with the CoO as the temperature falls below $T_N$ of the thin film.  The anisotropy field $\sim 1/\chi_{yy}$ increases substantially below about 200 K, and the peak in the dissipation marks $T_N$.  The dissipation represents the ability of even a small transverse field to overcome the weak interfacial coupling near $T_N$, and cause irreversible rotation of the magnetization.  Since no field cooling has been performed, a blocking temperature is not observed. 

Specific heat measurements\cite{Tang}, and previous studies using the anisotropic magnetoresistance\cite{Hunte} of CoO/Co bilayers suggest that a N\'{e}el temperature of 200K corresponds to a film thickness of about 1.5 nm $\sim$ 6 ML.  The thickness of the CoO in the present bilayers is not known, but this estimate is consistent with previous findings for CoO films grown by exposure to oxygen gas.\cite{Gruyters,Duo,Castro}.  The breadth of the peak seen in fig.(\ref{fig5}) is likely due to the nonuniform $T_N$ of a thin polycrystalline film.

\section{Conclusions}
Because the measurement of anisotropy fields is important in many aspects of research in ferromagnetism, it is essential to have a wide variety of measurement techniques that can satisfy the wide variety of practical constraints that are imposed by different experiments.  In UHV studies of thin films, it is often convenient to have a simple, accessible \textit{in situ} measurement technique that does not require a strong, static magnetic field.  In these cases, MOKE measurements of the magnetic susceptibility in a small a.c. field are a practical and accurate alternative to more well-established techniques.  The method is self-calibrating, independent of domain wall processes, but can also provide information about dissipative processes.  However, when a large static field is not applied, it is restricted to remanent geometries.  If a moderate d.c. transverse field could be applied, the method could be used to separate out various contributions to the total anisotropy by mapping the angular dependence of the anisotropy field.

A first demonstration of the method to Co/W(110) films shows that susceptibility measurements with MOKE provide anisotropies consistent with, and of similar accuracy to those measured using torque magnetometry.  The method can be simply adapted to measure the temperature dependence of the anisotropy field or its evolution as the film is grown.  The sensitivity of the method is demonstrated by measuring the anisotropy field created by the AFM/FM interface exchange coupling in CoO/Co/W(110) bilayers.  This allows the N\'{e}el temperature of a thin CoO film to be determined.


\begin{acknowledgments}
Financial support for this work  was provided by the Natural Sciences and Engineering Research Council of Canada.   A. B. acknowledges an Ontario Graduate Scholarship from the province of Ontario.
\end{acknowledgments}

\bibliography{anisotropy}

\end{document}